\documentclass[%
 aip,
 jcp,
citeautoscript,
 amsmath,amssymb,
 preprint,%
]{revtex4-2}
\usepackage[utf8]{inputenc}
\usepackage{textcomp}
\usepackage{graphicx}
\usepackage[version=4]{mhchem}
\usepackage{siunitx}
\DeclareSIUnit\au{a.u.}
\usepackage{amsmath}
\usepackage{upgreek}
\usepackage{hyperref}
\usepackage{amssymb}
\usepackage{tabularx}
\usepackage{booktabs}
\usepackage{epstopdf}
\AppendGraphicsExtensions{.tiff}
\usepackage{comment}
\usepackage{xcolor}
\allowdisplaybreaks

\colorlet{review}{red}

\makeatletter
\def\@email#1#2{%
 \endgroup
 \patchcmd{\titleblock@produce}
  {\frontmatter@RRAPformat}
  {\frontmatter@RRAPformat{\produce@RRAP{*#1\href{mailto:#2}{#2}}}\frontmatter@RRAPformat}
  {}{}
}%
\makeatother

\begin{document}


\title{On the Functional Dependence of Transition-Potential Coupled Cluster}

\author{Alexis Antoinette Ann Delgado}
\affiliation{Department of Chemistry, Southern Methodist University, Dallas, TX 75275, USA}

\author{Devin A. Matthews*}
\email{damatthews@smu.edu}
\affiliation{Department of Chemistry, Southern Methodist University, Dallas, TX 75275, USA}

\begin{abstract}

Orbital relaxation of the core region is a primary source of error in the computation of core ionization and core excitation energies. Recently, Transition-Potential Coupled Cluster (TP-CC) methods have been used to explicitly treat orbital relaxation using non-variational molecular orbitals determined by reoccupation of orbitals optimized for a fractional core occupation. The amount of fractional occupation is governed by parameter $\lambda$, and recommended values for accurate TP-CCSD and XTP-CCSD computations of carbon, nitrogen, oxygen, and fluorine 1s K-edges were previously determined. Herein, we explore the performance of a several density functionals for generating the fractionally occupied orbitals used in TP-CCSD. These functionals include HF, BP86, BH\&HLYP, B3LYP, M06-2X, and $\omega$B97m-V. The fractionally occupied orbitals computed across the various functionals were subsequently employed as the initial orbitals for our TP-CCSD calculations of organic K-edge x-ray absorption and photoelectron spectra. Regardless of the functional used to generate the fractionally occupied orbitals, the TP-CCSD calculations yield accurate and comparable core ionization energies, core excitation energies, and oscillator strengths.
\end{abstract}

\maketitle

X-ray spectroscopy is commonly employed to gain insight on electronic and molecular structure.\cite{Bokhoven2016, Bergmann2017, Kraus2018} These insights are vital for understanding molecular structure and behavior in depth. Simulated spectra are useful for unraveling complicated experimental spectra. It is well known that computed core ionization and core excitation energies are prone to error due to the relaxation of the core orbital. Traditionally, density functional theory (DFT) methods have been used for computing x-ray photoelectron (XPS) and x-ray absorption spectroscopy (XAS),\cite{Norman2018} either via linear response (i.e. TD-DFT)\cite{Gross1990} or through direct calculation of the core-hole state\cite{Hu1996,Triguero1998,Michelitsch2019,Triguero1999}.  Alternatively, transition-potential DFT (TP-DFT) overcomes the issue of orbital relaxation via the use of a fractionally populated reference state.\cite{Carbone2019} Equation-of-motion coupled cluster (EOM-CC) methods have also been used to compute XAS and XPS. EOM-CC generates excited states and ionized states from the correlated ground state via a response operator $\hat{R}$ that incorporates at least two-electron excitations,\cite{EOM-CC-1, EOM-CC-2, EOM-CC-3} or optionally triple or higher excitations in order to overcome correlation and core-hole relaxation errors.\cite{Matthews2020}

More recently, the most desirable traits of both the TP-DFT and EOM-CC methods were combined to form the transition-potential coupled cluster (TP-CC) method.\cite{Simons2021}  The TP-CCSD method counterbalances the overestimation of the core ionization and core excitation energies at the singles and doubles level, without the inclusion of triple excitations. This is because TP-CCSD accounts for the core orbital relaxation effects explicitly via orbitals optimized with a fractional core-hole occupation. These orbitals combine a partially destabilized ground state with a partially stabilized core-hole or core-excited state, leading to significant error cancellation. A parameter $\lambda$, which governs the fraction of core electrons removed from the core region, determines the exact point at which an error cancellation occurs between the two states.  This $\lambda$ parameter was previously optimized for generating fractionally occupied core-hole or core-excited orbitals at the B3LYP level of theory.\cite{Simons2022} The generated orbitals were then used to compute accurate TP-CCSD and XTP-CCSD 1s K-edge XAS and XPS for a small data set containing carbon, nitrogen, oxygen, and fluorine.\cite{Simons2022} In this Letter, we explore the dependence of TP-CCSD on the particular DFT functional used to determine the fractionally-occupied orbitals. In particular, we are interested in determining if a simple Hartree--Fock treatment (HF---that is, 100\% exact exchange and 0\% correlation) is adequate. This would simplify the incorporation of TP-CCSD into a wide range of quantum chemistry programs.

Our test set is comprised of 1s K-edge principal core ionization energies and four core excitation energies (for every 1s orbital) for \ce{H2O}, HCN, \ce{NH3}, CO, \ce{CH2}, \ce{CH4}, \ce{H2CNH}, \ce{H2NF}, \ce{H2CO}, \ce{H3CF}, \ce{H3COH}, HF, HNO, and HOF. Fractionally occupied core orbitals, required for the TP-CCSD calculations, were computed using the HF, BH\&HLYP\cite{Becke1988}, B3LYP\cite{Becke1993, Lee1988}, BP86\cite{Becke1988, Perdew1986}, M06-2X\cite{Zhao2007}, and $\omega$B97m-V\cite{Mardirossian2014} functionals, where the results from the B3LYP functional as well as the reference CVS-EOM-CCSDT values are taken from previous work.\cite{Simons2022} For the HF, BH\&HLYP, and BP86 functionals, the PSIXAS\cite{Ehlert2020} plugin in Psi4\cite{Smith2020} was used to generate fractionally occupied core orbitals. The $\lambda$ values we used were recommended in Table 2 of Ref.~\citenum{Simons2022}: carbon 1s ($\lambda$=0.350), nitrogen 1s ($\lambda$=0.375), oxygen 1s ($\lambda$=0.475), and flourine 1s ($\lambda$=0.425). Because the second derivatives for the M06-2X and $\omega$B97m-V functionals were not available in our version of Psi4 (1.3.2 release), we used Q-Chem\cite{Epifanovsky2021} to generate fractionally occupied core orbitals for these functionals. For the Q-Chem calculations the fractional occupation of the core orbital is limited to two decimal places, in this case the values used were: carbon 1s ($\lambda$=0.35), nitrogen 1s ($\lambda$=0.38), oxygen 1s ($\lambda$=0.48), and flourine 1s ($\lambda$=0.43). All ionization energies, excitation energies, and oscillator strengths were computed using TP-CCSD via the CFOUR program package.\cite{CFOUR} The aug-cc-pCVTZ basis set was used for all calculations excluding \ce{H2O} where the aug-cc-pCVQZ basis set was applied.

\begin{figure}
\includegraphics[width=0.7\textwidth]{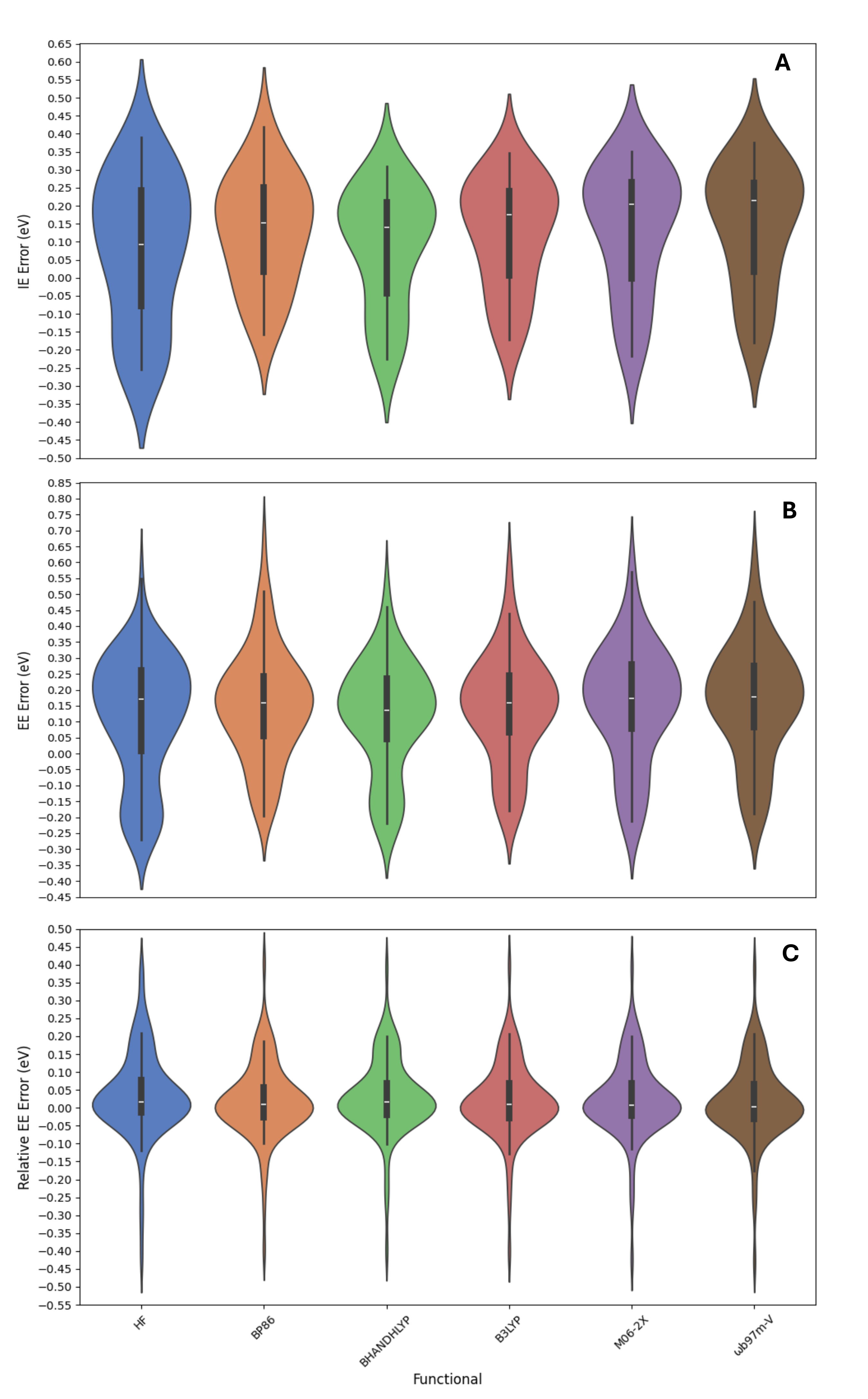}\caption{Error distributions for \textbf{A)} absolute (vertical) ionization energies, \textbf{B)}, absolute (vertical) excitation energies, and \textbf{C)}, relative excitation energies for orbitals obtained with various functionals. Shaded regions (violin plots) indicated the density of points while the box plots indicate the median and inter-quartile range, with whiskers indicating minimum and maximum errors, excluding statistical outliers.
\label{fig:energy_errors}}
\end{figure}

The distribution of absolute (vertical) IE errors from TP-CCSD is shown in Figure~\ref{fig:energy_errors}\textbf{A}. The error distributions for orbitals derived from BH\&HLYP, B3LYP, M06-2X, and \mbox{$\omega$B97m-V} are highly similar, featuring a signficant concentration of points around slightly positive errors ($\sim 0.2$~eV), and a smaller concentration of points extending to negative errors of about the same magnitude. For BP86 orbitals, the inter-quartile range (indicated by the shaded region of the box plot) is similar as for the other DFT functionals, but with errors shifted to the positive side by approximately 0.1~eV. The bimodality of the distribution is also somewhat less pronounced. The errors for Hartree--Fock orbitals are overall slightly larger. On the positive end, errors are similar as for BP86, but this is contrasted with more significant underestimation of the ``lower'' group of points. In total, errors for HF orbitals are increased by $\sim 20\%$ compared to DFT orbitals including any correlation component in the functional.

In contrast, the distributions of errors for computed absolute (vertical) excitation energies, as shown in Figure~\ref{fig:energy_errors}\textbf{B}, do not show any appreciable increase in error based on choice of orbitals. Here again, a bimodal distribution is observed, most strongly in the HF and BH\&HLYP results, and almost disappearing in BP86. While the underestimation of some points with HF is still evident, the effect is much smaller than for ionization energies. This likely points to a consistent cancellation of errors between correlation and/or relaxation effects on the core and virtual orbitals (or even the valence occupied orbitals, as these also play a role in the relaxation of the core hole). As before, orbitals derived from BH\&HLYP, B3LYP, M06-2X, and $\omega$B897m-V provide almost identical results, indicating a very weak effect of normally important features such as fraction of exact exchange, range separation, and non-local correlation. The obvious, though still numerically small, differences between these functionals and BP86 point to the correlation functional as the most important factor for this purpose. Despite these differences, the high degree of similarity between all six distributions supports the use of any self-consistent orbitals for TP-CCSD excitation energy calculations.

Additionally, we computed ``relative'' excitation energy errors by first shifting the computed TP-CCSD excitation spectra such that the TP-CCSD ionization energy aligns with the EOM-CCSDT benchmark value. The distributions of these errors are shown in Figure~\ref{fig:energy_errors}\textbf{C}. Here, the bulk of the errors fall between -0.1 and 0.1 eV, with a smaller number extending up to about 0.2 eV. A small number of outliers fall outside this range, notably the errors for the 3s Rydberg states of the HOF F 1s and HCN N 1s spectra. In general, errors are noticeably more concentrated around zero than in Figure~\ref{fig:energy_errors}\textbf{B}, and the range of errors is considerably compressed owing to significant error cancellation between the excitation and ionization energies. Interestingly, the error distribution for Hartree--Fock is extremely similar to that for all DFT methods, highlighting a consistent cancellation of errors originating both from the relaxation effect of the excited and ionized states, but also of errors arising from the orbitals themselves.

\begin{figure}
\includegraphics[width=0.7\textwidth]{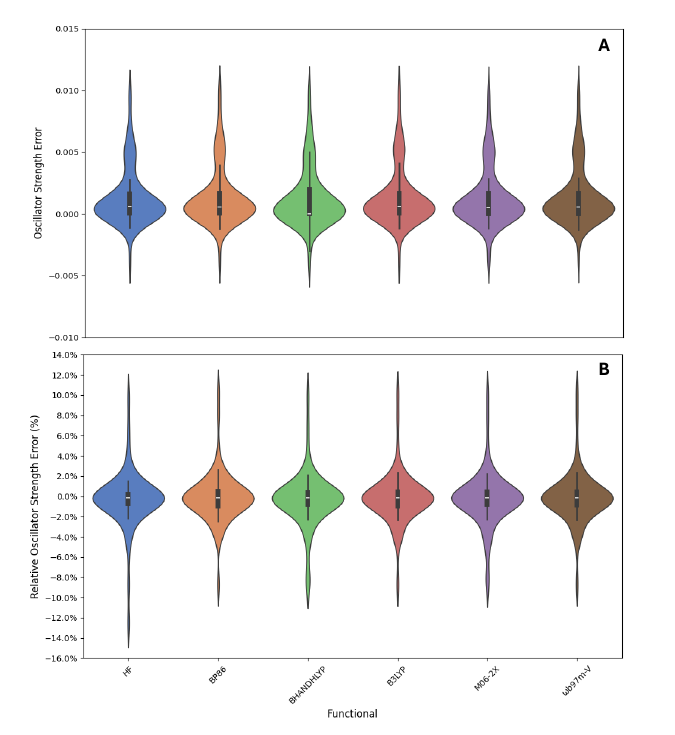}\caption{Error distributions for \textbf{A)} absolute oscillator strengths, \textbf{B)}, relative oscillator strengths for orbitals obtained with various functionals. Shaded regions (violin plots) indicated the density of points while the box plots indicate the median and inter-quartile range, with whiskers indicating minimum and maximum errors, excluding statistical outliers. \label{fig:os_errors}}
\end{figure}

The distributions of computed absolute oscillator strength errors are shown in Figure~\ref{fig:os_errors}\textbf{A}. The majority of errors fall below $\pm 0.003$, far below the typical intensity differences characteristic of strong XAS pre-edge and main edge features. Overall, TP-CCSD tends to over-estimate intensities, although the largest over-estimation is consistently of the most intense peaks. Thus, in Figure~\ref{fig:os_errors}\textbf{B}, normalization of intensities by setting the strength of the most intense peak in each spectrum to unity reveals \emph{relative} errors of only 2--3\% in most cases. The largest relative deviations occur consistently for very weak transitions, but even then are limited to approximately 10\%. The choice of orbitals has essentially no effect on the computed oscillator strengths, particularly for the absolute oscillator strengths and for transitions of medium to high intensity.

In summary, our calculations of TP-CCSD K-edge ionization energies, excitation energies, and associated oscillator strengths using orbitals derived from fractional occupation calculations with a range of DFT and Hartree--Fock functionals show that the role of the specific functional is minimal, with the fractional occupation effect itself dominating the accuracy of TP-CCSD calculations. The one exception is a slight deterioration in the accuracy of vertical ionization energies for uncorrelated Hartree--Fock orbitals, with errors relative to EOM-CCSDT increasing by approximately 20\%. In describing the internal structure of the absorption pre-edge region (measured via relative excitation energies and oscillator strengths), the variation due to choice of functional is almost undetectable. These findings suggest that a simple Hartree--Fock treatment of the fractionally occupied orbitals in TP-CCSD is wholly sufficient, which greatly simplifies the application of TP-CCSD and it's incorporation into a wide array of quantum chemistry programs.

\begin{acknowledgments}
This work was supported by the US National Science Foundation under grant CHE-2143725 and by the Department of Energy under grant DE-SC0022893. Computational resources for this research were provided by SMU’s O’Donnell Data Science and Research Computing Institute.
\end{acknowledgments}

\section*{Conflict of Interest Statement}

The authors have no conflicts to disclose.

\section*{Supplementary Information}

Available supplemental information files:

\begin{itemize}
\item \texttt{SupportingInformation.xlsx}: All excitation energies (eV), oscillator strengths, CCSDT excitation energy references (eV), CCSDT oscillator strength references,	excitation energy errors (eV), oscillator strength errors, relative excitation energy errors (eV), relative oscillator strength errors, and relative oscillator strength percent errors computed in this work.
\end{itemize}

\section*{Data Availability}

The data that support the findings of this study are available within the article and its supplementary material.

\bibliography{paper}

\end{document}